\documentclass[aps,prl,twocolumn,balancelastpage,showpacs,superscriptaddress]{revtex4-1}

\usepackage{graphicx}
\usepackage[letterpaper,total={7in,9.5in},top=0.75in,left=0.75in]{geometry}

\begin{document}

\onecolumngrid
\title{Realization of SU(2)$\times$SU(6) Fermi System}
\author{Shintaro Taie}
\altaffiliation{To whom correspondence should be addressed. E-mail: taie@scphys.kyoto-u.ac.jp}
\affiliation{Department of Physics, Graduate School of Science, Kyoto University, Japan 606-8502}
\author{Yosuke Takasu}
\affiliation{Department of Physics, Graduate School of Science, Kyoto University, Japan 606-8502}
\author{Seiji Sugawa}
\affiliation{Department of Physics, Graduate School of Science, Kyoto University, Japan 606-8502}
\author{Rekishu Yamazaki}
\affiliation{Department of Physics, Graduate School of Science, Kyoto University, Japan 606-8502}
\affiliation{CREST, JST, 4-1-8 Honcho Kawaguchi, Saitama 332-0012, Japan}
\author{Takuya Tsujimoto}
\affiliation{Department of Physics, Graduate School of Science, Kyoto University, Japan 606-8502}
\author{Ryo Murakami}
\affiliation{Department of Physics, Graduate School of Science, Kyoto University, Japan 606-8502}
\author{Yoshiro Takahashi}
\affiliation{Department of Physics, Graduate School of Science, Kyoto University, Japan 606-8502}
\affiliation{CREST, JST, 4-1-8 Honcho Kawaguchi, Saitama 332-0012, Japan}
\date{\today}

\begin{abstract}
We report the realization of a novel degenerate Fermi mixture with an SU(2)$\times$SU(6) symmetry in a cold atomic gas.
We successfully cool the mixture of the two fermionic isotopes of ytterbium ${}^{171}\text{Yb}$ with the nuclear spin $I=1/2$ and ${}^{173}\text{Yb}$ with $I=5/2$
below the Fermi temperature $T_{\rm F}$ as  0.46$T_{\rm F}$ for ${}^{171}\text{Yb}$ and 0.54$T_{\rm F}$ for ${}^{173}\text{Yb}$.
The same scattering lengths for different spin components make this mixture featured with the novel SU(2)$\times$SU(6) symmetry.
The nuclear spin components are separately imaged by exploiting an optical Stern-Gerlach effect.
In addition, the mixture is loaded into a 3D optical lattice to implement the SU(2)$\times$SU(6) Hubbard model. 
This mixture will open the door to the study of novel quantum phases such as a spinor Bardeen-Cooper-Schrieffer-like fermionic superfluid.
\end{abstract}
\maketitle
\begin{figure}[b]
\hbox to\hsize{
\begin{minipage}{\textwidth}
\centering
\includegraphics[width=175mm]{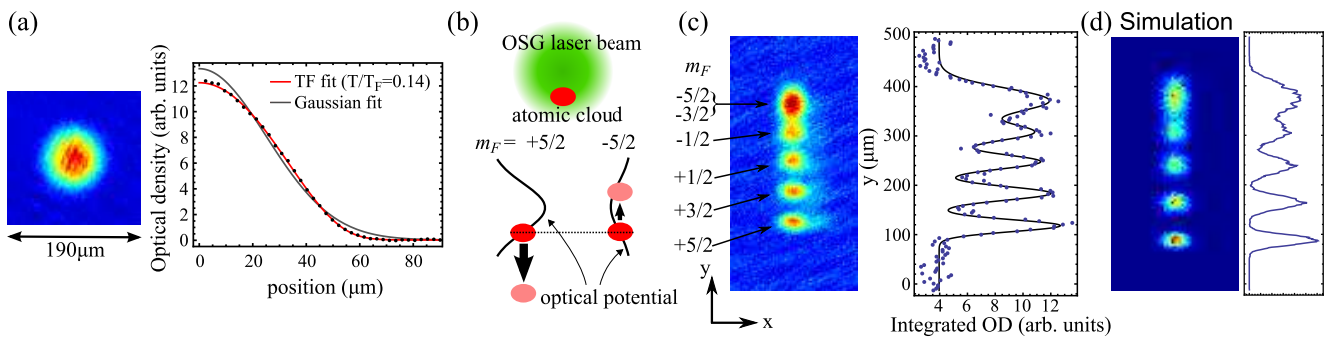}
\caption{(Color online) Optical Stern-Gerlach separation of nuclear spins. (a) Time-of-flight image of a degenerate Fermi gas of ${}^{173}\text{Yb}$
without the OSG separation. 
The image is taken after 12 ms ballistic expansion. The azimuthally averaged distribution is also shown on the right hand side.
The temperature of $0.14T_{\rm F}$ is determined from the Thomas-Fermi fit (red line). The observed distribution clearly
deviates from the classical Gaussian shape, indicated by the gray line.
(b) Schematic view of an OSG effect. The OSG beam has the waist of about $100\ \mu$m, the pulse duration of 2.5 ms
and the beam power of $4$ mW.
The atoms in the $m_F=+5/2$ state is pushed downward in the figure, whereas the $m_F=-5/2$ upward.
(c) Optical Stern-Gerlach separation of spin components in the Fermi gas of $^{173}\text{Yb}$.
The expansion time is 8 ms.
Integration of the images along the horizontal axis are also shown on the right hand side.
(d) The simulated distribution under the current experimental condition is shown.
\label{image1}
}
\end{minipage}
\hss}
\end{figure}
While an SU(2) symmetry is ubiquitous in nature, realization of an SU($N$) symmetry with $N>2$ in condensed matter physics is a rather special case.
A system with higher symmetry is expected to show novel behaviors in both qualitative and quantitative ways.
One remarkable example is the Kondo effect in quantum dots, 
in which the SU(4) symmetry due to orbital degeneracy considerably increases the Kondo temperature, compared to the SU(2) case \cite{Kondo}.
The SU(4) symmetry is also approximately realized in graphene, and relevant quantum Hall magnetism is discussed \cite{QHE}.
Ultracold atomic gases seem to be good candidates for exploring the physics of SU($N$) because of their variety of spin degrees of freedom
and high controllability.
However, it is difficult, though not impossible \cite{SUNhon}, to realize enlarged spin symmetries with alkali atoms because of their complicated hyperfine structures.
This difficulty can be overcome by using fermionic isotopes of alkaline-earth-metal-like atoms in which the absence of electronic spin in the ground states
decouples their nuclear spin from collision processes \cite{SUNcaz}. In this case, the scattering lengths for any combination of spin components are the same,
resulting in an SU($2I+1$) symmetry for nuclear spin $I$ \cite{SUNcaz}.
Recently, multicomponent fermions with higher symmetry attract much attention and the presence of rich quantum phases are predicted \cite{SUNhon,SUNcaz,SUNher,SUNgor}.
In particular, a system of two-orbit SU($N$) symmetry is intensively discussed in Ref. \cite{SUNgor}.  
\begin{figure}[b]
\vspace{80mm}
\end{figure}

In this Letter, we report the realization of a two-species Fermi-Fermi degenerate gas mixture with a novel SU(2)$\times$SU(6) spin symmetry.
This is attained by applying an all-optical evaporative cooling method to two ytterbium (Yb) fermionic isotopes of 
${}^{171}\text{Yb}$ with the nuclear spin $I=1/2$ and ${}^{173}\text{Yb}$ with $I=5/2$.
We successfully cool down both isotopes with full spin degrees of freedom,
below the Fermi temperature as 0.46$T_{\rm F}$ for
${}^{171}\text{Yb}$ and 0.53$T_{\rm F}$ for ${}^{173}\text{Yb}$. 
While the spin-polarized two-species Fermi-Fermi mixture \cite{LiK} is attractive to study many unexplored quantum phenomena \cite{FFphase},
it is natural to expect rich quantum phases with the Fermi-Fermi mixture with the SU(2)$\times$SU(6) spin symmetry.
As a first step, we load the mixture into a 3D optical lattice to implement the SU(2)$\times$ SU(6) Hubbard model where a variety of quantum phases has been recently discussed \cite{SUNgor}.
Another intriguing example is the possibility of a spinor Bardeen-Cooper-Schrieffer (BCS) -like fermionic superfluidity
discussed in Ref. \cite{spinorBCS} where an interesting similarity between the heteronuclear superfluids and spinor BEC at the mean-field level has been pointed out.

\begin{figure}[bth]
\includegraphics[width=75mm]{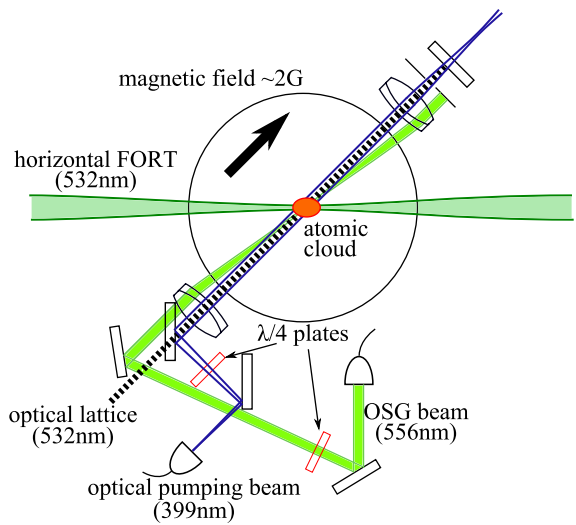}
\caption{
(Color online) Experimental setup. Laser beams for the OSG separation, optical pumping, and an optical lattice
are shown. The OSG laser beam, whose wavelength is close to that of the lattice, is slightly tilted with respect to others.
This does not cause any serious errors in the observed spin population.
}
\label{image2}
\end{figure}

The experimental procedure for preparing two-species mixture of Yb isotopes is described in detail in Ref. \cite{Ybmixture}.
Below we briefly summarize the method.
Figure \ref{image2} shows the schematic view of our experimental setup.
Yb atoms are loaded into a magneto-optical trap (MOT) with the ${}^{1}$S$_0 \leftrightarrow {}^{3}$P$_{1}$ transition
($\lambda = 556 \ \text{nm}$) after the deceleration by a Zeeman-slower using the ${}^{1}$S$_0 \leftrightarrow {}^{1}$P$_{1}$
transition ($\lambda = 399 \ \text{nm}$).
In order to trap the two isotopes simultaneously in the MOT, we use two independent MOT light beams.
After collecting about $10^7$ ${}^{173}\text{Yb}$ atoms in about 12 second, the frequency of the Zeeman slower beam is
changed to load  ${}^{171}\text{Yb}$ atoms for about 5 second.
By optimizing the MOT laser detuning and intensity, we obtain about $10^7$ atoms for both isotopes, before the sympathetic evaporative cooling.
Next, the two isotopes are loaded into a crossed far-off resonant trap (FORT) with $532$ $\text{nm}$ light.
Typically we obtain $2\times 10^5\ {}^{171}\text{Yb}$ and $8\times 10^5\ {}^{173}\text{Yb}$ atoms in the FORT.
The sympathetic evaporative cooling is performed by continuously decreasing the FORT trap depth. 
From our previous photoassociation study \cite{2PA}, the small $s$-wave scattering scattering length of $-0.15$ nm is known for ${}^{171}\text{Yb}$,
whereas that for ${}^{173}\text{Yb}$ is as large as $10.55$ nm. 
We can employ efficient sympathetic cooling of ${}^{171}\text{Yb}$ by an evaporatively cooled ${}^{173}\text{Yb}$
through the large inter-species scattering length of $-30.6$ nm. 

Before we describe the results for the two-species mixture, we show the results for single-species Fermi gas of
${}^{173}{\rm Yb}$ with an SU(6) symmetry to illustrate the nuclear spin selective imaging with an Optical Stern-Gerlach (OSG) effect.
The spin selective imaging or detection is crucially important in many studies of Fermi gases with spin degrees of freedom.
The separate imaging of each nuclear spin components is, however, very difficult with a usual Stern-Gerlach effect due to the small nuclear
magnetic moments.
Here in this work we successfully overcome this difficulty by exploiting the OSG effect produced by
an off-resonant circularly polarized laser beam \cite{OSG}.
It is noted that the spin-dependent light shift is the origin of the fictitious magnetic field
considered here \cite{FictBfield}.
Figure \ref{image1} (a) shows absorption images of pure ${}^{173}{\rm Yb}$ degenerate gases with 6 components
without the OSG separation.
The fit to the Thomas-Fermi profile yields the temperature of $0.14T_{\rm F}$ and the number of atoms of $5.0 \times 10^4$.
The Fermi temperature is given by
\begin{equation}
T_{\text{F}}=\left(6N/s\right)^{1/3}\hbar\omega/k_{\rm B}, \label{eq1}
\end{equation}
where $N$ is atom number, $s=6$ for ${}^{173}{\rm Yb}$ is the number of degenerate states, $\hbar$ is the Planck constant over $2\pi$, $\omega$ is
mean trap frequency, and $k_{\rm B}$ is the Boltzmann constant.
While the achievement of quantum degeneracy of ${}^{173}{\rm Yb}$ has already been reported \cite{173DFG},
the spin population was not investigated.

Figure \ref{image1} (b) shows a schematic view of the OSG experiment.
The OSG beam is focused just above the atom cloud with the waist of about $100\ \mu$m to provide the atoms
an dipole force due to the potential gradient.
In this measurement, the pulse with the duration of $2.5$ ms, the beam power of $4$ mW, and the detuning of about $+1$ GHz
with respect to the $ {}^{1}S_0 (F_g=5/2) \leftrightarrow {}^{3}P_{1} (F_e=7/2)$ transition of ${}^{173}\text{Yb}$ is used.
Figure \ref{image1} (c) shows the separately observed images of spin components of ${}^{173}\text{Yb}$.
The integrated distributions of the image along the horizontal axis are also shown on the right hand side. 
Figure \ref{image1} (d) shows the simulated distributions under the present experimental condition and assumption of no spin polarization.
The overall feature of the observed distributions can be reproduced. The distortion of the initially isotropic momentum distribution
is caused by the non-uniform intensity gradient of the OSG beam.
In order to image ${}^{173}\text{Yb}$ atoms with $m_F=-5/2$ and $m_F=-3/2$ states separately, 
we repeat the measurement with an opposite sense of circular polarization for the OSG beam.
We find that the relative imbalance of the spin population is smaller than 5\%, which shows almost no spin polarizations.
The demonstrated technique is very useful for investigating the novel magnetism induced for a system with high spin symmetry. 
\begin{figure}[t]
\includegraphics[width=75mm]{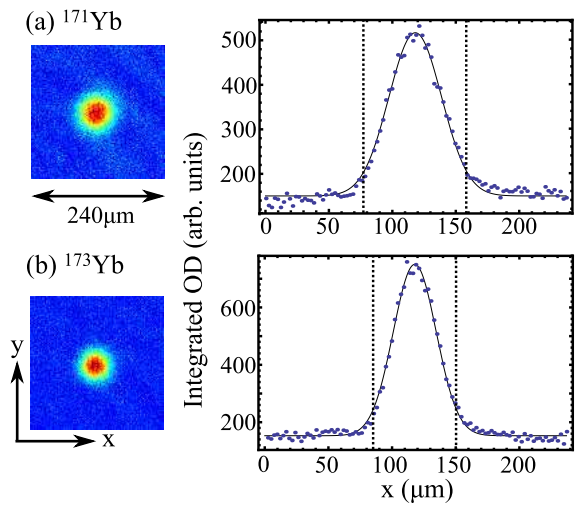}
\caption{(Color online) Time of flight images of the quantum degenerate Fermi-Fermi mixture with spin degrees of freedom, (a) for two-spin mixture of ${}^{171}\text{Yb}$ and (b) for 6-spin mixture of ${}^{173}\text{Yb}$. 
The expansion time is 9 ms for ${}^{171}\text{Yb}$ and 8 ms for ${}^{173}\text{Yb}$. 
The density distributions integrated over the vertical direction $y$ are also shown on the right hand side. 
From the Gaussian fits we can estimate the temperatures T=95 nK and T=87 nK for ${}^{171}\text{Yb}$ and ${}^{173}\text{Yb}$, respectively.
The dotted lines correspond to the Fermi velocities $v_{\rm F}=\sqrt{2k_{\rm B}T_{\rm F}/m}$ for each isotope.
The images are averaged over 5 independent measurements.
}
\label{image3}
\end{figure}

Next we describe the experimental results for two-species Fermi-Fermi mixture.
Figure \ref{image3} shows absorption images obtained after the final stage of the evaporative cooling.
Starting from the $2\times 10^5\ {}^{171}\text{Yb}$ and $8\times 10^5\ {}^{173}\text{Yb}$ atoms in the FORT,
about $12$ second evaporation results in the coldest temperatures of less than 100 nK.
It is noted that the image of each isotope is taken using the two independent probe beam with a sequential measurement
for the same sample.
We fit the measured momentum distribution with a Gaussian distribution and obtain
the atom numbers of ${}^{171}\text{Yb}$ and ${}^{173}\text{Yb}$ are $8.0\times10^3$  and $1.1\times10^4$, respectively.
The temperatures are $95$ $\text{nK}$ for ${}^{171}\text{Yb}$ and $87$ $\text{nK}$ for ${}^{173}\text{Yb}$, and we estimate
$T/T_{\text{F}}$ to be $0.46$ for ${}^{171}\text{Yb}$, and $0.54$ for ${}^{173}\text{Yb}$, respectively \cite{temperature}.
In this temperature regime, the fits with the Fermi-Dirac distribution give almost the same value for $T/T_{\text{F}}$ as the Gaussian distributions.
\begin{figure}[t]
\includegraphics[width=80mm]{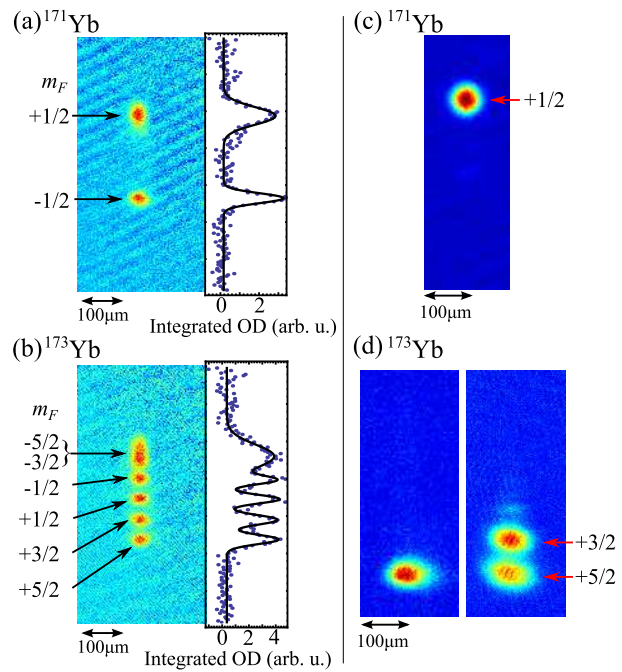}
\caption{(Color online) Left: Optical Stern-Gerlach separation of spin components in the ${}^{171}\text{Yb}-{}^{173}\text{Yb}$ quantum degenerate mixture
(a) for ${}^{171}\text{Yb}$ and (b) for ${}^{173}\text{Yb}$.
The expansion time is 6 ms and the images are averaged over 8 independent measurements.
Integrations of the images along the horizontal axis are also shown on the right hand side.\\
Right: Spin manipulation by optical pumping, applied to single-species samples. (c) The OSG separation is applied after optical pumping to the $m_F=+1/2$
state of ${}^{171}$Yb. (d) For ${}^{173}$Yb, optical pumping allows to prepare either a single-component gas in the $m_F=+5/2$ state (left)
or a two-component mixture of the $m_F=+3/2$ state and the $m_F=+5/2$ state (right).
}
\label{image4}
\end{figure}

The OSG separation is also applied to the mixture gas of ${}^{171}\text{Yb}$ and ${}^{173}\text{Yb}$.
The same detuning of OSG laser light as used for pure ${}^{173}{\rm Yb}$ sample is also applicable for simultaneously
separate the nuclear spin components of both isotopes. Again, we confirm that the atoms are almost equally distributed over all nuclear spin states
for both ${}^{171}\text{Yb}$ and ${}^{173}\text{Yb}$ (see Figs. \ref{image4} (a) and (b)), indicating that the system has the SU(2)$\times$SU(6) symmetry.

It is worth noting that by applying optical pumping to ${}^{173}\rm{ Yb}$ via the ${}^1S_0 (F_g=5/2) \leftrightarrow {}^3P_1(F_e=3/2)$ transition,
we can prepare an almost equal mixture of $m_F=+3/2$ and $m_F=+5/2$ states. In this case, an SU(2)$\times$SU(2) symmetry is realized.
In addition, optical pumping also enables to create a spin polarized mixture of ${}^{171}\text{Yb}$ and ${}^{173}\text{Yb}$.
Here, the ${}^1S_0 (F_g=1/2) \leftrightarrow {}^1P_1(F_e=1/2)$ transition for ${}^{171}\text{Yb}$
and the ${}^1S_0 (F_g=5/2) \leftrightarrow {}^1P_1(F_e=5/2)$ transition for ${}^{173}\text{Yb}$ are used for optical pumping.
The suppression of three-body losses in two-component Fermi gases allows us to cool the sample down to lower temperature,
$0.33T_{\rm F}$ for ${}^{171}\text{Yb}$ and $0.30T_{\rm F}$ for ${}^{173}\text{Yb}$.
The spin distribution after optical pumping is examined by the OSG separation and shown in Figs. \ref{image4} (c) and (d).

\begin{figure}[bt]
\includegraphics[width=80mm]{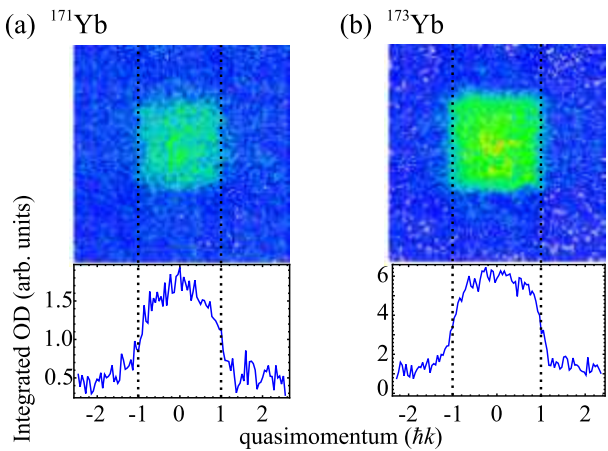}
\caption{(Color online) Quasimomentum distribution of (a) ${}^{171}\text{Yb}$ and (b) ${}^{173}\text{Yb}$ in the SU(2)$\times$SU(6) two-species mixture in an optical lattice.
The density distributions integrated along  the vertical direction are also shown below.
The atom numbers are $0.4 \times 10^4$ for ${}^{171}\text{Yb}$ and $1.5 \times 10^4$ for ${}^{173}\text{Yb}$, respectively.
The images are taken after linear ramping down of the lattice in 0.5ms,
followed by a ballistic expansion of (a) 12ms and (b) 13ms. The dotted lines indicate the domain of the 1st Brillouin zone, which equals twice the recoil
momentum $\hbar k$.
}
\label{image5}
\end{figure}

Finally, the mixture is successfully loaded into a 3D optical lattice to implement the SU(2)$\times$SU(6) Hubbard model. 
A variety of quantum phases in such a system is discussed \cite{SUNgor}.
The 3D lattice is formed with three orthogonal standing waves with a lattice constant of $266$ nm. 
Figure \ref{image5} shows the quasimomentum distribution (a) for ${}^{171}\text{Yb}$ and (b) for ${}^{173}\text{Yb}$ for the lattice height of $10 E_r$ measured by a band-mapping technique. 
Here $E_r$ is the recoil energy and is about 200 nK in this experiment.
One can see that the momentum spreads over the entire first Brillouin zone for both fermions,
which is typical of an insulating regime \cite{Koehl}.
This behavior is well reproduced by a theoretical calculation with no interaction taken into consideration.
The strong interaction effect is revealed by our recent observation of Bloch oscillations in a 3D optical lattice, which is beyond the scope of this paper and will be discussed elsewhere.

In conclusion, we demonstrate the successful realization of two-species Fermi-Fermi degenerate gas mixture of the fermionic isotopes of
${}^{171}\text{Yb}$ with $I=1/2$ and ${}^{173}\text{Yb}$ with $I=5/2$ with spin degrees of freedom.
The nuclear spin components for each fermion are separately imaged by exploiting an optical Stern-Gerlach effect. 
The coldest temperatures achieved are 0.46$T_{\rm F}$ and 0.54$T_{\rm F}$ for ${}^{171}\text{Yb}$  for ${}^{173}\text{Yb}$, respectively.
The mixture is successfully loaded into a 3D optical lattice to implement the SU(2)$\times$SU(6) Hubbard model.   
Various kinds of novel quantum phases could be studied by this Fermi-Fermi mixture with the spin degrees of freedom. 
Especially interesting is to investigate a spinor BCS-like fermionic superfluid with the technique of the optical Feshbach resonance.
Due to the large negative inter-species scattering length of $-30.6 $nm, an efficient optical Feshbach resonance effect is expected \cite{OFRexperiment,OFRtheory}. 
In fact, quite recently we find several ${}^{171}\text{Yb}-{}^{173}\text{Yb}$ heteronuclear photoassociation resonances which can be used for the optical Feshbach resonance experiments.

We acknowledge M. A. Cazalilla for first pointing out the SU(6) symmetry for ${}^{173}\text{Yb}$ and importance of separate spin detection,
and S. Uetake and T. Fukuhara for their experimental help.
This work was supported by the Grant-in-Aid for Scientific Research of JSPS (No. 18204035, 21102005C01 (Quantum Cybernetics),
21104513A03 (DYCE), 22684022),
GCOE Program ``The Next Generation of Physics, Spun from Universality and Emergence'' from MEXT of Japan,
FIRST, and Matsuo Foundation.
ST and SS acknowledge supports from JSPS.


\begin{thebibliography}{99}
\bibitem{temperature}
The strong attractive interaction between ${}^{171}\text{Yb}$ and ${}^{173}\text{Yb}$
can affect the width of the measured momentum distribution.
We estimate the increase in the trap frequency due to the mean field potential
to be $20-30$\%, resulting in the reduction of the actual $T/T_{\rm F}$.
On the contrary, the conversion of the negative interaction energy into the kinetic energy
during the expansion time causes the underestimation of the temperature by $\sim 10\%$.
Therefore we can expect that the actual $T/T_{\rm F}$ of the sample is $\sim 15$\% lower than that reported in the text.
\end{thebibliography}
\end{document}